\input epsf
\documentstyle[preprint,eqsecnum,aps]{revtex}
\newcommand\LQCD{{\Lambda_{\rm QCD}}}

\count255=\time\divide\count255 by 60 \xdef\hourmin{\number\count255}
  \multiply\count255 by-60\advance\count255 by\time
  \xdef\hourmin{\hourmin:\ifnum\count255<10 0\fi\the\count255}

\begin{document}

%\twocolumn[]

\draft
\preprint{\vbox{\hbox{UCSD/PTH 99--02} \hbox{JLAB-THY-99-05}}}

\title{$B^+ \to D^{*+}_s \gamma$ and $B^+ \to D^{*+} \gamma$ as Probes
of $V_{ub}$}

\author{Benjam\'{\i}n Grinstein\footnote{bgrinstein@ucsd.edu}}

\address{Department of Physics,
University of California at San Diego, La Jolla, CA 92093}

\author{Richard F. Lebed\footnote{lebed@jlab.org}}

\address{Jefferson Lab, 12000 Jefferson Avenue, Newport News, VA
23606}

\date{February, 1999}

\maketitle
\tightenlines

\widetext

\begin{abstract}
The decays $B^+ \to D^{*+}_s \gamma$ and $B^+ \to D^{*+} \gamma$ can
be used for an extraction of $|V_{ub}|$.  When the $b$ and $c$ quarks
are nearly degenerate  the rate for these modes can be determined in
terms of other observed rates, namely
$B\overline{B}$ mixing and $D^* \to D \gamma$ decay. To this end we
introduce  a novel
application of heavy quark and flavor symmetries. Although somewhat
unrealistic, this limit provides us with a first estimate of these rates. 
\end{abstract}


\narrowtext

\section{Introduction}

        The extraction of the Cabibbo-Koba\-yashi-Maskawa (CKM) mixing
matrix element $V_{ub}$, theoretically and experimentally, stands out
as one of the prominent challenges of particle physics.  Both its
magnitude (through decay rates) and its phase relative to other CKM
elements (through CP violation) are of considerable interest.
Nevertheless, its determination remains elusive, ultimately because
the CKM matrix describes the mixing of quarks, whereas of course only
hadrons are observed.  In the case of $V_{cb}$, much of this
impediment is overcome by the application of the heavy quark effective
theory (HQET) \cite{HQET,SV}, which in particular relates the strong
interaction matrix elements of systems with $b$ and $c$ quarks,
eliminating much of the strong interaction uncertainty.  However, the
$u$ quark is by no means heavy, and it is notoriously difficult to
separate its CKM and strong interaction couplings.

        This is true even for the experimentally clean semileptonic
modes, $\overline{B} \to \pi \ell \bar \nu$ and $\overline{B} \to \rho
\ell \bar \nu$, although dispersive bounds on the shapes of the form
factors help to restrict such strong interaction uncertainties
\cite{disp}.  Inclusive semileptonic rates for $\overline{B} \to X_u
\ell \bar \nu$ are theoretically under better control but are 
plagued, since $|V_{cb}| \gg |V_{ub}|$, by the
preponderance of $\overline{B} \to X_c \ell \bar \nu$ everywhere in
kinematic space except in the endpoint region of maximal lepton
energy, where a $c$ quark cannot kinematically be produced; however,
in this small region it has proved necessary to include hadronic model
dependence 
\cite{incl}. 
A technique involving invariant hadronic mass spectra for invariant
mass below $m_D$ is under better theoretical control but requires
neutrino reconstruction \cite{hadrmass}.

        Hadronic modes, containing everywhere strong interaction
uncertainties, are even more problematic.  Even if these difficulties
could be tamed, one would still encounter the mixing of different weak
topologies.  For example, $\overline{B}^0 \to \pi^+ K^-$ may proceed
through either $b \to u s \bar u$ ($\propto V^{\mbox{ }}_{ub}
V_{us}^*$) or a penguin $b \to s g \to s \bar u u$
($\propto V^{\mbox{ }}_{tb} V_{ts}^*$); the presence of spectator
quarks complicates the analysis.

        One may probe $V_{ub}$ using the decays $B^+ \to D^{*+}_s
\gamma$ ($\bar b u \to c \bar s \gamma$) and $B^+ \to D^{*+} \gamma$
($\bar b u \to c \bar d \gamma$)\cite{vssp}, collectively $B^+ \to
D^{*+}_{(s)} \gamma$.  Even though the second decay is
Cabibbo-suppressed compared to the first, the lower reconstruction
efficiency associated with $D^{*+}_s$ compared to $D^{*+}$ makes both
processes worth studying.  The hard, monochromatic photon in these
decays, 2.22 and 2.26 GeV, respectively, provides a distinctive
experimental signature.

        As shown in Fig.~1, the invariant amplitude for these decays
at $O(G_F^1)$
consists of only one weak topology, because four flavor quantum
numbers change ($\Delta B = -1$, $\Delta C = +1$, $\Delta (S,D) = +1$,
$\Delta U = -1$), thus fixing the quark couplings to the $W$
boson.\footnote{At $O(G_F^2)$, the nontrivial diagrams include the
di-penguin $\bar b \to \bar s g$, $u g \to c$ and a box diagram with
the internal quarks crossed.}  
The photon may couple to any of the quark lines, although one
expects the dominant contributions from
radiation emitted by the light $u$ or $\bar s$. 
The
diagram with the photon emitted from the $W$ is suppressed, of
$O(G_F^2)$.
On the other hand, since gluons may
pair-produce quarks, these decays may be sensitive to multiparticle
intermediate states, for example $B^+ \to D^0 K^+ \to D_s^{*+}
\gamma$~\cite{NI}, but the weak topology remains unchanged.

        In this paper we study a theoretical limit in which the decays
$B^+ \to D^{*+}_{(s)} \gamma$ are calculable from first principles in
terms of other measured quantities. We assume both the $b$ and $c$
quarks are heavy and nearly degenerate, $\LQCD\ll m_b$, $\LQCD\ll m_c$ and
$\delta m\equiv m_b- m_c \alt \LQCD$. We do not expect this limit to
be a good approximation to reality. Our calculation is a starting point for
further investigations on controlled approximations of this rate.   
\begin{figure} \label{fig1}
  \begin{centering}
        \def\epsfsize#1#2{0.75#2}
        \hfil\epsfbox{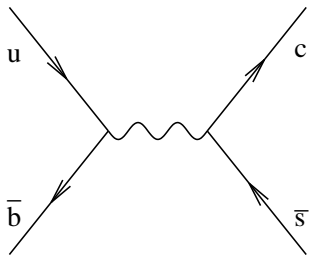}\hfil\hfil

\medskip

\caption{The unique weak diagram topology at $O(G_F^1)$ for $B^+ \to
D^{*+}_s \gamma$.  The diagram for $B^+ \to D^{*+} \gamma$ simply
changes $\bar s$ to $\bar d$.}

  \end{centering}
\end{figure}

\section{Formalism}

        The dominant contributions to the amplitude for $B^+ \to
D_{(s)}^{*+}\gamma$ are the long distance processes $B^+ \to D^+_{(s)}\to
D_{(s)}^{*+}\gamma$ and $B^+ \to B^{*+} \gamma \to D_{(s)}^{*+}\gamma$. The
premise of this calculation is that the matrix elements of $B^+ \to
D^+_{(s)}$, $D^+_{(s)}\to D_{(s)}^{*+}\gamma$, and $B^+\to B^{*+}\gamma$ can be
related by heavy quark and $SU(3)$ flavor symmetries to the rates for
$B \overline{B}$ mixing and $D_{(s)}^*\to D_{(s)}\gamma$.
Alternately, we expect that lattice calculations similar to those that
determine the $B \overline{B}$ mixing matrix element\cite{BaBar} can
be employed to compute that of $B^+\to D_{(s)}^+$ directly.

        For the initial calculation presented here, we work in the
generally-low (GL) velocity limit
\begin{equation}
\delta m\equiv m_b - m_c \alt \LQCD \ll m_b ,
\end{equation}
which means that the four-velocity of each heavy particle in the
process is assumed to remain constant to lowest order.  
This limit is more restrictive than the slow-velocity (SV) limit
considered by Shifman and Voloshin \cite{SV}, 
$\LQCD \ll \delta m  \ll m_b  $.
To indicate
the quality of this assumption, we point out that the value of
$\gamma_{D^{*}_{(s)}} = v_B \cdot v_{D^*_{(s)}}$ is 1.45 for the
strange case and 1.50 for the nonstrange case.  On the other hand,
letting $m_B = m_b + \Lambda -3\lambda^2/4m_b$, 
$m_{D^*_{(s)}} = m_c + \Lambda +\lambda^2/4m_c$,
one finds that $\gamma_{D^{*}_{(s)}}$ is parametrically small,
\begin{eqnarray}
\gamma^{\mbox{ }}_{D^{*}_{(s)}} & = & \frac{m_B^2 +
m_{D^*_{(s)}}^2}{2m_B m_{D^*_{(s)}}} \nonumber \\ & = & 1 +
\frac{1}{2m_b^2} \delta m^2
+ O \left( \frac{1}{m_b^3} \right)
. 
\end{eqnarray}

	Heavy quark symmetries may be used to relate the matrix
element for $B\to D_{(s)}$ to the one for $B\to\overline B$ only if
the velocity of the $D_{(s)}$ in the $B$ rest frame is parametrically
small, say $O(\Lambda/m_b)$ or $O(\delta m/m_b)$.
It is therefore sufficient to assume
the weaker SV limit for this part of our argument. However, in order
to relate the photon emission processes to the real decays $B^*\to
B\gamma$ and $D^*\to D\gamma$, it is necessary to impose the GL
condition to ensure softness of the photon, as we now argue. 

	The matrix element of the electromagnetic current $j_\mu$
between  vector ($D^*$ or $B^*$) and pseudoscalar ($D$ or $B$) states
is characterized by a single Lorentz invariant form factor
\begin{equation}
\langle {\vec p}\;', \vec\varepsilon \, |j_\mu(0)| \, \vec p \, \rangle
=-ig(q^2)\epsilon_{\mu\nu\lambda\sigma}\varepsilon^{*\nu} 
p^{\prime\lambda}p^\sigma,
\end{equation}
where $q=(p-p')$. The form factor at $q^2=0$ governs the on-shell decay of
the vector to the pseudoscalar, which is characterized by a transition
magnetic moment, $g(0)=e\mu$ [see below, Eq.~(\ref{radrate})].
However, implicit in this definition is that the mesons are on-shell
states; if the transition occurs with one of the states being virtual,
as is the case here, then the $q^2$ argument of the form factor must
be computed accordingly.  The relevant scale over which a hadronic
form factor changes is $\LQCD$.  For example, in the
non-relativistic potential quark model of mesons, the scale that
dictates the behavior of the form factor $g$ is the constituent quark
mass $\sim\LQCD$. In $B^+\to D^{*+}_{(s)}\gamma$ the form factor
appears with argument $q^2\simeq -\delta m^2$, and in
the GL limit one may approximate $g(-\delta m^2)\simeq g(0)=e\mu$.

        The formalism of HQET combined with chiral symmetry, relevant
to the physics of mesons containing heavy quarks, was developed in
Refs.~\cite{Wise}.  It was extended to radiative processes, such as
$D^* \to D \gamma$, in Ref.~\cite{polar}, while the Lagrangian for
heavy quark-heavy antiquark mixing appeared in \cite{GJMSW}.  We use
ingredients from all of these works, but for brevity include only
notation relevant to the present process. While convenient, the
formalism is not really a necessary framework for our calculation. In
fact, light mesons enter diagrams only at higher order in the
chiral expansion.  By stating our calculation in the language of this
effective Lagrangian we are setting the stage for further
investigations of, for example, the effects of the finite strange
quark mass.

        Since chiral symmetry is one of the ingredients of the
Lagrangian, we begin by including the octet of pseudo-Nambu--Goldstone
bosons in the usual nonlinear form $\xi$:
\begin{equation}
\xi = \exp ( i \Pi/f ),
\end{equation}
where
\begin{equation}
\Pi = \left( 
\begin{array}{ccc} {{\textstyle 1} \over \sqrt{\textstyle
2}} \pi^0 + {{\textstyle 1} \over \sqrt{\textstyle 6}} \eta & \pi^+ &
K^+ \\ \pi^- & - {{\textstyle 1} \over \sqrt{\textstyle 2}} \pi^0 +
{{\textstyle 1} \over \sqrt{\textstyle 6}} \eta & K^0 \\ K^- &
\overline{K}^0 & - {\sqrt{\textstyle 2 \over \textstyle 3}} \eta
\end{array}
\right) ,
\end{equation}
with $f \simeq 131$ MeV.  Under the chiral SU(3)$_{\rm L}$ $\times$
SU(3)$_{\rm R}$ transformation ($L,R$),
\begin{equation}
\xi \mapsto L \xi U^\dagger = U \xi R^\dagger,
\end{equation}
with $U$ implicitly defined by the equality of these two forms.

        The ground state vector $P^*_{Aa}$ and pseudoscalar $P^{\mbox{
}}_{Aa}$ fields that destroy mesons of heavy quark flavor $A$ and
light antiquark flavor $a$ are incorporated into the 4 $\times$ 4
bispinor
\begin{eqnarray}
H_{Aa} & = & \frac{( 1 + v \hskip-0.5em / )}{2} \left( P^{*\mu}_{Aa}
\gamma_\mu - P^{\mbox{ }}_{Aa} \gamma_5 \right) , \nonumber \\
\overline{H}^{Aa} & = & \gamma^0 H_{Aa}^\dagger \gamma^0 .
\end{eqnarray}
Under SU(4) heavy quark spin-flavor symmetry transformations $S$ and
SU(3)$_{\rm L}$ $\times$ SU(3)$_{\rm R}$ transformations $U$, $H_a$
transforms as
\begin{eqnarray}
H_{Aa} & \mapsto & S_A{}^B \, H_{Bb} \, U^{\dagger \, b}{}_a, \nonumber \\
\overline{H}^{Aa} & \mapsto & U^{\, a}{}_b \, \overline{H}^{Bb} \,
S^\dagger{}_B{}^A .
\end{eqnarray}
Defining the chiral covariant derivative and axial current by
\begin{eqnarray}
D^\mu_{ab} & = & \delta_{ab} \partial^\mu - V^\mu_{ab} \nonumber \\ &
= & \delta_{ab} \partial^\mu - \frac 1 2 \left( \xi^\dagger
\partial^\mu \xi + \xi \partial^\mu \xi^\dagger \right)_{ab} =
\delta_{ab} \partial^\mu + O(\Pi^2), \nonumber \\ A^\mu_{ab} & = &
\frac{i}{2} \left( \xi^\dagger \partial^\mu \xi - \xi \partial^\mu
\xi^\dagger \right)_{ab} = -\frac 1 f \partial^\mu \Pi_{ab} +
O(\Pi^3),
\end{eqnarray}
one finds that $D\xi$, $D\xi^\dagger$, $DH$, and $D\overline{H}$ have
precisely the same transformation properties under chiral symmetry as
their underlying fields, and
\begin{equation}
\left( A^\mu \right)^a{}_b \mapsto U^{\, a}{}_c \,
\left( A^\mu \right)^c{}_d \, U^{\dagger \, d}{}_b .
\end{equation}
Thus, one obtains the heavy meson Lagrangian invariant under chiral
and heavy quark symmetry with the minimum number of derivatives,
\begin{equation} \label{Lag}
{\cal L} = - {\rm Tr} \, \overline{H}^{Aa} i v \cdot D^b{}_a H_{Ab} +
g \, {\rm Tr} \, \overline{H}^{Aa} H_{Ab} \hskip 0.25em /
\hskip-0.75em A^b{}_a \gamma_5.
\end{equation}
It should be noted that the kinetic term is canonically normalized
when the field $H$ has 2, rather than the usual $2M$, particles per
unit volume.

	Since lowest-order HQET integrates out heavy antiparticle
degrees of freedom, it is necessary to include such fields explicitly
when they can appear in the asymptotic states.  One
defines\cite{GJMSW}
\begin{eqnarray}
H^{\overline{A}a} & = & \left( P^{*\overline{A}a}_\mu \gamma^\mu -
P^{\overline{A}a} \gamma_5 \right) \frac {( 1 - v \hskip-0.5em / )}{2}
, \nonumber \\ \overline{H}_{\overline{A}a} & = & \gamma^0
H^{\dagger\overline{A}a} \gamma^0 .
\end{eqnarray}
For example, while the field $P^*_{Aa}$ destroys a vector meson of
flavor content $A \bar a$, the field $P^{*\overline{A}a}$ destroys one
of flavor content $\overline{A} a$; the two are related by a chosen
charge conjugation convention\cite{GJMSW}.  One demands that $H_{Aa}$
and $H^{\overline{A}a}$, or $\overline{H}^{Aa}$ and
$\overline{H}_{\overline{A}a}$, have the same transformation
properties under heavy quark symmetry and chiral transformations.
Then the construction of the Lagrangian for mesons with heavy
antiflavors is straightforward.

        As for direct mixing between the two sectors, the standard
model operator that destroys quarks of flavors $\overline{A},a$ and
creates quarks of flavors $B, \bar b$, given charge constraints $ -
Q_A + Q_a = Q_B - Q_b$, is
\begin{equation} \label{SM}
{\cal O}^{ab,\overline{A}B} = \overline{A} \gamma^\mu (1-\gamma_5) a
\cdot \overline{B} \gamma_\mu (1-\gamma_5) b .
\end{equation}
For $B\overline{B}$ mixing, one sees that $A$ and $B$ are $b$ quarks,
and $a$ and $b$ are $d$ quarks; in the present case, $A$, $B$, $a$,
and $b$ are $b$, $c$, $u$, and ($d$ or $s$) quarks, respectively.  In
the limit that HQET and SU(3) are good symmetries, both processes have
the same strong matrix element.  In Ref.~\cite{GJMSW} it was shown
that (\ref{SM}) matches in the symmetry limit of the effective theory
onto
\begin{equation} \label{mix}
{\cal O}^{ab,\overline{A}B}  =  4\beta C \left[ \left( \xi P^{*
  \dagger\mu} \right)^{\overline{A}a} \left( \xi P^*_\mu \right)^{Bb}
  + \left( \xi P^\dagger \right)^{\overline{A}a} \left( \xi P
  \right)^{Bb} \right] + \cdots ,
\end{equation}
where $C$ is the color vacuum saturation coefficient, 1 for $B
\overline{B}$ mixing and 3/8 for $BD_{(s)}$.

	Calculation of the process also requires understanding the
coupling of heavy mesons to photons.  Minimal substitution for the
kinetic term in Eq.~(\ref{Lag}) provides no coupling to real
transverse photons at lowest order, 
while minimal substitution applied to the second term in
(\ref{Lag}) requires including at least one pion for a nonvanishing
term.  The unique
lowest-order electromagnetic term\cite{polar} is the transition
magnetic moment operator
\begin{equation} \label{mag}
\delta {\cal L} = \frac{1}{4} e \mu^{\mbox{ }}_{Aa} {\rm Tr} \,
\overline{H}^{Aa} H_{Aa} \sigma^{\mu\nu} {\cal F}^{\rm EM}_{\mu\nu} +
\mbox{\rm charge conjugate} .
\end{equation}

\section{Calculation}

	The photon coupling in Eq.~(\ref{mag}) represents the
lowest-order heavy-to-heavy electromagnetic transition.  It does not
couple pseudoscalars to each other, but provides for the decay of
vector mesons:
\begin{equation}
\label{radrate}
\Gamma (P^*_{Aa} \to P_{Aa} \gamma) = \frac 1 3 \alpha \, |\mu_{Aa}|^2
E_\gamma^3 .
\end{equation}
To lowest order in HQET and SU(3), one may write\footnote{In Eq.~(14)
of Ref.~\cite{polar}, $\mu$ is labeled $\beta$.} $\mu^{\mbox{ }}_{Aa}
= Q_a \mu$.  For $P^*_{Aa} = D^{*+}$, the current experimental
bound\cite{PDG} of $\Gamma (D^{*+}) < 0.131$ MeV and electromagnetic
branching fraction $< 3.2\%$, lead to $|\mu| < 2.5$ GeV$^{-1}$.

	The calculation of the rate is then straightforward in the
combined HQET and chiral symmetry limit, and consists of the two
diagrams depicted in Fig.~2.  We use the GL limit to compute the
amplitude, but retain full mass dependence in the phase space
calculation. The result is
\begin{equation}
\label{resultrate}
\Gamma (B^+ \to D^{*+}_s \gamma) = \frac{G_F^2}{64} \left| V_{ub}^*
V^{\mbox{ }}_{cs} \right|^2 (3 \beta /2)^2 \alpha |\mu|^2
\frac{m_{D_s^*}}{m_B^4}(m_B-m_{D_s^*})(m_B+m_{D_s^*})^3 \ ,
\end{equation}
with $V_{cs} \to V_{cd}$ and $m_{D_s}^* \to m_D^*$ for $B^+ \to D^{*+}
\gamma$.

\begin{figure} \label{fig2}
  \begin{centering}
        \def\epsfsize#1#2{3.4#2}
        \hfil\epsfbox{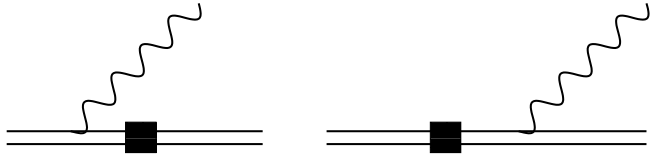}\hfil\hfil

\medskip

\caption{The two leading diagrams for $B^+ \to D^{*+}_{(s)} \gamma$
that occur in the heavy quark and chiral limits.  The double line
indicates a heavy meson, one containing a $\bar b$ ($c$) quark on the
left (right) of the flavor-changing vertex, indicated by a box.}

  \end{centering}
\end{figure}

	The effective Lagrangian mixing parameter $\beta$ in
(\ref{mix}) is related to familiar quantities of $B\overline{B}$
mixing via
\begin{equation}
\frac{1}{m_B}\left< \overline{B}^0 \left| {\cal O}^{dd,\overline{B}B}
\right| B^0 \right> = \frac 8 3 f_B^2 m_B B_B = 4\beta .
\end{equation}
A full treatment requires including renormalization point dependence
in $B_B$, but we neglect this here for simplicity.  Now there are two
obvious directions of analysis.  One is to eliminate $\beta$ using its
calculated value from lattice simulations.  Using \cite{BaBar} $f_B =
170 \pm 35$ MeV and $B_B = 0.98 \pm 0.06$ (renormalization point 2
GeV), and $m_B = 5.279$ GeV, $\tau_B = 1.6 \cdot 10^{-12}$ s, one
finds $4\beta = 0.40 \pm 0.17$ GeV$^3$, and
\begin{eqnarray}
\lefteqn{{\rm Br} (B^+ \to D^{*+}_s \gamma) \simeq 2 \cdot 10^{-7}
\cdot \left( \frac{B_B}{0.98} \right)^2} & & \nonumber \\ & & \cdot
\left( \frac{\Gamma (D^{*+})}{0.131 \mbox{ {\rm MeV}}}
\right) \cdot \left( \frac{{\rm Br} (D^{*+} \to D^+ \gamma)}{3.2\%}
\right) \cdot \left| \frac{V_{ub}^* V^{\mbox{ }}_{cs}}{3 \cdot
10^{-3}} \right|^2 ,
\end{eqnarray}
with $V_{cs} \to V_{cd}$ for $B^+ \to D^{*+} \gamma$, in which case
the coefficient (taking $|V_{cd}| = 0.22$) is $7 \cdot 10^{-9}$.  The
other direction for analysis eliminates $\beta$ as a systematic strong
interaction uncertainty between this process and $B\overline{B}$
mixing.  In particular, $\Delta m_B \propto G_F^2 |V^{\mbox{ }}_{tb}
V_{td}^*|^2 \beta$, which means\footnote{Note that
$\Gamma(B^+ \to D^{*+}_{(s)} \gamma) \propto G_F^2 \beta^2$, while
$\Delta m_B \propto G_F^2 \beta$, due to the different weak topologies.}
that eliminating $\beta$ leads to
values for the ratio $|V^*_{ub} V^{\mbox{ }}_{cs}|^2 / |V^{\mbox{
}}_{tb} V_{td}^*|^4$, which is proportional to $(\rho^2 +
\eta^2)/[(1-\rho)^2 + \eta^2]^2$ in the usual Wolfenstein space.
Curves of this family tend to intersect the $\rho$ axis with vertical
slope, which is useful since the current experimentally allowed region
in $\rho$-$\eta$ space is broad in the $\rho$ direction.

\section{Prospects}

	The inclusion of HQET-violating, chiral symmetry-violating,
and short-distance corrections is straightforward, and this program
should certainly be carried out.  Indeed, many of the necessary
corrections already appear in the literature, including corrections to
$\mu_{Aa}$\cite{polar}, meson decay constants\cite{GJMSW,BG},
$B\overline{B}$ mixing\cite{GJMSW}, and a number of other
SU(3)-symmetry corrections\cite{GJMSW,FG}.  However, a number of other
diagrams remain to be computed, because the inclusion of the photon
changes the allowed quantum numbers of intermediate states.  For
example, there are new chiral loop diagrams for $BD_{(s)}$ mixing
compared to $B\overline{B}$ mixing where a single pion emerges
directly from the mixing vertex.

	More urgent and less tractable is understanding of the
radiative form factor $g(q^2)$ at realistic momenta. Taking $g(q^2)
\to g(0) = e\mu$ introduces
large uncertainties into our estimate for the decay
rate. Alternatively, one may use only the less extreme SV limit to
relate the mixing amplitudes and express the rate in terms of the
corresponding form factors. This is done by replacing $e\mu$ in
Eq.~(\ref{resultrate}) by $g_B(q^2)+g_D(q^2)$ at $q^2=-\delta m^2$,
which can be used a starting
point for a calculation based on estimates using hadronic models for
the form factors.

	We expect that the best determination of $V_{ub}$ will still
require careful analysis of a number different decay modes; however,
the decays $B^+ \to D^{*+}_{(s)} \gamma$ provide an important
additional handle on the problem.

\vskip1in
{\it Acknowledgments} This work is supported by the Department of
Energy under contract Nos.\ DOE-FG03-97ER40546 and DE-AC05-84ER40150.


\begin{thebibliography}{199}
\bibitem{vssp} The viability of this mode was suggested to us by the
fine CLEO and BABAR experimentalists Vivek Sharma and Soeren Prell.
\end{thebibliography}
\end{document}